\documentclass[conference,a4paper]{APSIPA2025}
\usepackage{amsmath}
\usepackage{graphicx}
\usepackage{multirow}
\usepackage{threeparttable}
\usepackage[backend=biber,style=ieee,]{biblatex}
\usepackage{setspace}

\usepackage{geometry}
\usepackage{hhline}
\usepackage{tabularx}
\newcommand{\Hline}{\noalign{\hrule height 0.8pt}}
\usepackage{fancyhdr}
\usepackage{comment}

\usepackage[labelformat=simple]{subcaption}

\makeatletter
\renewcommand\p@subfigure{\thefigure\,}
\makeatother
\fancypagestyle{firststyle}{
  \fancyhf{}
  \fancyhead[C]{2025 Asia Pacific Signal and Information Processing Association Annual Summit and Conference (APSIPA ASC)}
}

\begin{document}

\title{Human-CLAP: Human-perception-based\\contrastive language--audio pretraining}

\author{
\authorblockN{
Taisei Takano\authorrefmark{1},
Yuki Okamoto\authorrefmark{1},
Yusuke Kanamori\authorrefmark{1},
Yuki Saito\authorrefmark{1},
Ryotaro Nagase\authorrefmark{2},
Hiroshi Saruwatari\authorrefmark{1}
}

\authorblockA{
\authorrefmark{1}
The University of Tokyo, Japan \\
}

\authorblockA{
\authorrefmark{2}
Ritsumeikan University, Japan\\
}
}

\maketitle
\thispagestyle{firststyle}
\pagestyle{fancy}

\begin{abstract}
Contrastive language--audio pretraining (CLAP) is widely used for audio generation and recognition tasks.
For example, CLAPScore, which utilizes the similarity of CLAP embeddings, has been a major metric for the evaluation of the relevance between audio and text in text-to-audio.
However, the relationship between CLAPScore and human subjective evaluation scores is still unclarified.
We show that CLAPScore has a low correlation with human subjective evaluation scores.
Additionally, we propose a human-perception-based CLAP called Human-CLAP by training a contrastive language-audio model using the subjective evaluation score.
In our experiments, the results indicate that our Human-CLAP improved the Spearman's rank correlation coefficient (SRCC) between the CLAPScore and the subjective evaluation scores by more than 0.17 compared with the conventional CLAP.
\end{abstract}

\section{Introduction}
Various tasks focusing on audio--text relationships, such as text-to-audio (TTA)~\cite{liu2023audioldm,10112585,kreuk2023audiogen}, language-queried audio source separation (LASS)~\cite{lass, 10752098}, and automated audio captioning (AAC)~\cite{Drossos_2017_waspaa, shah2023_t6b, greeshma2023_t6a}, have been developed.
Constructing text representations of audio is a common effective approach across such audio--text tasks, making audio more interpretable for humans~\cite{wang2023_t6b}.
Representing audio as text descriptions improves the searchability of audio data and allows more people, including those with hearing difficulties, to enjoy audio contents.

CLAP~\cite{laionclap2023, elizalde2023clap} has been proposed as a foundation model for connecting audio and text by contrastive learning.
By learning the correspondence of audio--text pairs, CLAP creates a shared embedding space between audio and text, enabling wide applications to audio recognition and generation tasks~\cite{pmlr-v202-huang23i}.
Moreover, CLAP is used not only for these tasks, but also for the evaluation of relevance between audio and text in TTA~\cite{liu2024audioldm,majumder2024tango,ghosal2023text, 10731578, ijcai2024p502}.
Concretely, the distance between the audio and text embedding vectors obtained by CLAP can be interpreted as a score that quantifies the degree of relevance between the audio and the text, which is called CLAPScore.
Although CLAPScore is used for the objective evaluation in TTA tasks~\cite{liu2024audioldm,majumder2024tango}, the relationship between CLAPScore and human subjective evaluation scores has not been investigated.
Considering that TTA aims to generate sounds humans hear, it is necessary to investigate how well the CLAPScore aligns with human subjective evaluation scores.
In addition, although CLAP is trained with the assumption that the audio--text pairs in the dataset match exactly, the audio--text pairs also include noisy data, such as text that does not contain all the content in the audio.
Therefore, there is a possibility that the distance between audio and text will be close in the embedding space of CLAP, even for inappropriate audio--text pairs.
It is also possible to collect cleaner audio--text pairs for model training, but data collection is extremely costly.

In this paper, we first investigate the relationship between CLAPScore and subjective evaluation scores, focusing on the relevance between audio and text (REL).
We also propose a human-perception-based CLAP called Human-CLAP by fine-tuning a pretrained CLAP model using a small number of subjective scores, consisting of approximately 1/320th the training data of conventional CLAP, as an approximation of the ideal relevance of an audio-text pair.
By using scores that evaluate the degree of relevance of text and audio as determined by humans, we aim to construct a CLAP model that matches human perception.

In summary, our contributions are as follows:
\begin{itemize}
    \item We perform a subjective evaluation of the relevance between texts and audio samples.
    By comparing CLAPScore with collected subjective evaluation scores, we show that CLAPScore has a low correlation with subjective scores.
    \item We propose a human-perception-based CLAP called ``Human-CLAP'', which is trained using subjective scores.
    We experimentally show that the CLAPScore calculated by Human-CLAP has a higher correlation with subjective scores than the conventional CLAP.
\end{itemize}
\section{Relationship between CLAPScore and Human Perception}\label{sec:clap_subj}
Although the distance between CLAP embedding vectors is interpreted as the audio--text relevance, its alignment with human perception is still unclarified.
We examine how well CLAP matches human perception by investigating the correlation between CLAPScore and subjective evaluation scores.

\subsection{CLAPScore}
CLAPScore is an evaluation metric for the relevance of texts and audio samples in TTA.
CLAPScore utilizes CLAP~\cite{elizalde2023clap} embeddings and is defined as
\begin{equation}\label{eq:clapscore}
    \textsf{CLAPScore}=\max\left(\frac{\textbf{e}^{\textsf{audio}}\cdot \textbf{e}^{\textsf{text}}}{\|\textbf{e}^{\textsf{audio}}\| \|\textbf{e}^{\textsf{text}}\|},0\right),
\end{equation}
where $\textbf{e}^{\textsf{text}}$ and $\textbf{e}^{\textsf{audio}}$ denote the CLAP embedding of text and audio, respectively~\cite{liu2024audioldm}.
Although CLAPScore is sometimes calculated to take values between $-1$ and $1$, in this paper, we defined negative similarity scores as $0$ to compare correlation with human evaluation scores.
This definition follows the precedence set by the conventional work on CLIPScore~\cite{hessel-etal-2021-clipscore}.

\subsection{Dataset of subjective scores for audio--text pairs}\label{subsec:collection}
We used the RELATE dataset~\cite{kanamori_INTERSPEECH2025}, which contains subjective evaluation scores of the relevance between audio--text pairs.
The RELATE dataset includes natural audio samples from AudioCaps \cite{kim2019audiocaps} and synthesized audio samples generated using AudioLDM \cite{liu2023audioldm}, AudioLDM2 \cite{liu2024audioldm}, Tango \cite{ghosal2023text}, and Tango2 \cite{majumder2024tango}.
\begin{figure}[t]
    \centering
    \includegraphics[width=0.45\textwidth]{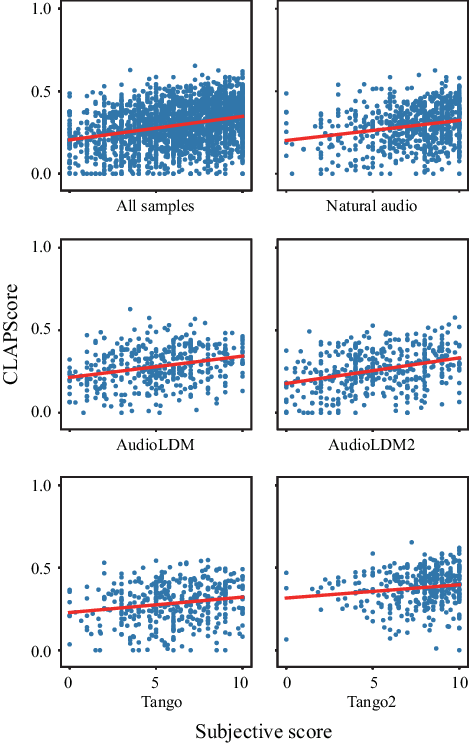}
    \caption{Relationship between CLAPScore and subjective scores (Model: LAION CLAP)}
    \vspace{-8pt}
    \label{fig:laion_test}
\end{figure}
In the RELATE dataset, audio--text pairs were presented to listeners, who were then asked, ``How well does the text description match the audio?''
Listeners rated each audio--text pair on an 11-point scale, where 0 indicated no match and 10 indicated a perfect match.
To suppress individual differences when rating, the corpus developer provided the following guidance when asking the question: 
\begin{description}
    \item[ 0 -] Does not match at all
    \item[ 2 -] Has significant discrepancies
    \item[ 5 -] Has several minor discrepancies
    \item[ 8 -] Has a few minor discrepancies
    \item[10 -] Matches exactly
\end{description}


To select high-quality listeners, we conducted a screening of listeners using anchor samples in the listening test.
The RELATE dataset provides three anchor samples, which were mismatched audio--text pairs, in each listening test.
We removed listeners who scored higher than 2 on average for anchor samples.
This screening method is based on the subjective evaluation conducted by Choi et al. in DCASE 2023 Challenge Task 7~\cite{choi:hal-04178960}.

Each audio--text pair was evaluated by four listeners on average.
After the screening, 2,383 audio--text pairs from the RELATE training set and 2,405 audio--text pairs from the RELATE test set were obtained.

%
\subsection{Correlation between CLAPScore and Subjective Evaluation Scores}
We investigated the relationship between CLAPScore and subjective evaluation score using 2,405 audio--text pairs from test data collected in Section~\ref{subsec:collection}.
Two types of CLAP model were used, models by Wu et al. (LAION CLAP)~\cite{laionclap2023} and Elizalde et al. (MS CLAP)~\cite{elizalde2023clap}.
To evaluate the correlation between CLAPScore and subjective evaluation scores, we used Spearman's rank correlation coefficient (SRCC), linear correlation coefficient (LCC), and Kendall's rank correlation coefficient (KTAU).

The CLAPScore calculated with LAION CLAP had an SRCC of $0.280$ and an LCC of $0.294$.
Similarly, those calculated with MS CLAP had an SRCC of $0.278$ and an LCC of $0.296$.
From these results, CLAPScore has a low correlation with subjective evaluation scores.
Such a low correlation is also visualized in the scatter plot of the relationship between CLAPScore and subjective score shown in Fig.~\ref{fig:laion_test}.
These results indicate that CLAPScore is not necessarily suitable for an objective evaluation metric in TTA.
\begin{figure}[t!]
\centering
\includegraphics[scale=0.85]{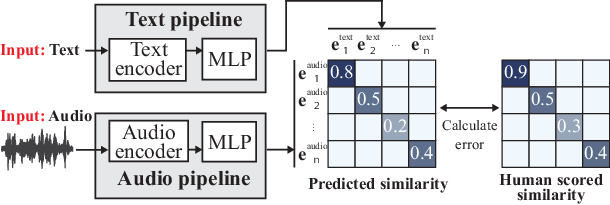}
\caption{Overview of Human-CLAP}
\label{fig:proposed}
\end{figure}

\section{Proposed Method}
\subsection{Framework and Motivation}
We propose Human-CLAP, which utilizes human subjective scores for training the CLAP model.
Figure~\ref{fig:proposed} shows the overview of our proposed method.
The conventional CLAP was trained to maximize the cosine similarity of embedding vectors for paired text and audio, while minimizing it for unpaired ones.
However, this approach heavily relies on the quality of the text labels in the dataset.
Even though some low-quality text labels might be included, the conventional method equally maximizes the similarity of each audio--text pair.
On the other hand, we expect the CLAP model to train the relationship between audio and text on the basis of human perception using subjective evaluation scores as target scores.
When using it as a target score, we rescaled the subjective score to the range of 0 to 1.

\subsection{Loss Function}
Our Human-CLAP model is trained to minimize a loss function that combines contrastive learning and regression approaches.
Regression losses such as the mean square error (MSE) and mean absolute error (MAE) are often used to train models that predict subjective evaluation scores for speech--text pairs or audio--text pairs~\cite{tjandra2025aes, baba2024utmosv2, lo19_interspeech}.
In this study, we used the regression loss between the target score $a_i\in[0,1]$ and the predicted cosine similarity $y_i$.
We utilize MSE and MAE to calculate such regression losses.
MSE is calculated as
\begin{equation}\label{eq:MSE}
\begin{aligned}
L_{\rm MSE} &= \frac{1}{N}\sum_{i=1}^N \left( a_i-y_i\right)^2,\\
\end{aligned}
\end{equation}
\begin{equation}\label{eq:y_n}
\begin{aligned}
y_i &= 
\mathsf{ReLU}\left(\frac{\textbf{e}_i^{\textsf{text}}\cdot\textbf{e}_i^{\textsf{audio}}}{\|\textbf{e}_i^{\textsf{text}}\|\|\textbf{e}_i^{\textsf{audio}}\|}\right), \\
\end{aligned}
\end{equation}
where $a_i\in [0,1]$ is the rescaled subjective score of the \textit{i}th audio--text pair in a batch, $\textbf{e}_i^{\textsf{text}}$ and $\textbf{e}_i^{\textsf{audio}}$ denote the embeddings of \textit{i}th data in batch, $\mathsf{ReLU}(\cdot)$ denotes the rectified linear unit, and $N$ denotes the batch size.
Similarly to MSE, we calculated MAE as
\begin{equation}\label{eq:MAE}
\begin{aligned}
L_{\rm MAE} &= \frac{1}{N}\sum_{i=1}^N \left|a_i-y_i\right|. \\
\end{aligned}
\end{equation}

We also trained the model with the contrastive learning approach performed in the conventional CLAP model~\cite{elizalde2023clap, laionclap2023}.
The conventional CLAP is trained by using symmetric cross entropy (SCE) loss~\cite{laionclap2023, DBLP:conf/icml/RadfordKHRGASAM21} to construct its audio--text embedding space.
We propose the weight SCE (wSCE) loss by adding weights based on the subjective score to the SCE loss:
\begin{equation}\label{eq:wSCE}
\begin{aligned}
L_{\rm wSCE}=&-\frac{1}{2N} \sum_{i=1}^N a_i \left(\log\frac{\exp(\textbf{e}_i^{\textsf{text}}\cdot\textbf{e}_i^{\textsf{audio}}/\tau)}{\sum_{j=1}^N \exp(\textbf{e}_i^{\textsf{text}}\cdot\textbf{e}_j^{\textsf{audio}}/\tau)} \right.\\
&\qquad + \left. \log \frac{\exp(\textbf{e}_i^{\textsf{audio}}\cdot\textbf{e}_i^{\textsf{text}}/\tau)}{\sum_{j=1}^N \exp(\textbf{e}_i^{\textsf{audio}}\cdot\textbf{e}_j^{\textsf{text}}/\tau)}\right),
\end{aligned}
\end{equation}
where $a_i$ denotes the rescaled subjective score of \textit{i}th audio--text pair and $\tau$ is a learnable temperature parameter for scaling the loss.
The SCE loss minimizes the distance between relevant audio--text embeddings and maximizes the distance between unpaired embeddings.
By weighting the SCE loss with subjective scores, we believe that the wSCE loss function will construct a better embedding space, reflecting human perception.
\begin{table}[tb]
\centering
\caption{Evaluation results of CLAPScore and subjective evaluation scores}
\label{tab:eval-proposed-eusipco}
\resizebox{\linewidth}{!}{%
\begin{tabular}{l|cccc}
\Hline
\multicolumn{1}{l|}{Model} & SRCC $\uparrow$ & LCC $\uparrow$ & KTAU $\uparrow$ & MSE $\downarrow$ \\
\hline
{\bf Human-CLAP (ours)}\\
\quad wSCE + MSE & 0.447 & 0.471 & 0.313 & 0.059 \\
\quad wSCE + MAE & \textbf{0.457} & \textbf{0.481} & \textbf{0.320} & 0.057 \\
\quad wSCE & 0.383 & 0.410 & 0.265 & 0.063  \\
\quad MSE & 0.439 & 0.462 & 0.307 & 0.056 \\
\quad MAE & 0.453 & 0.472 & 0.317 &\textbf{0.051}  \\

\hline
{\bf Baseline}\\
\quad LAION CLAP & 0.280 & 0.294 & 0.192 & 0.068 \\
\quad MS CLAP & 0.278 & 0.296 & 0.192 & 0.078 \\
\Hline
\end{tabular}
}
\end{table}
%
%
%
\begin{figure*}[t!]
\centering
\includegraphics[scale=1.0]{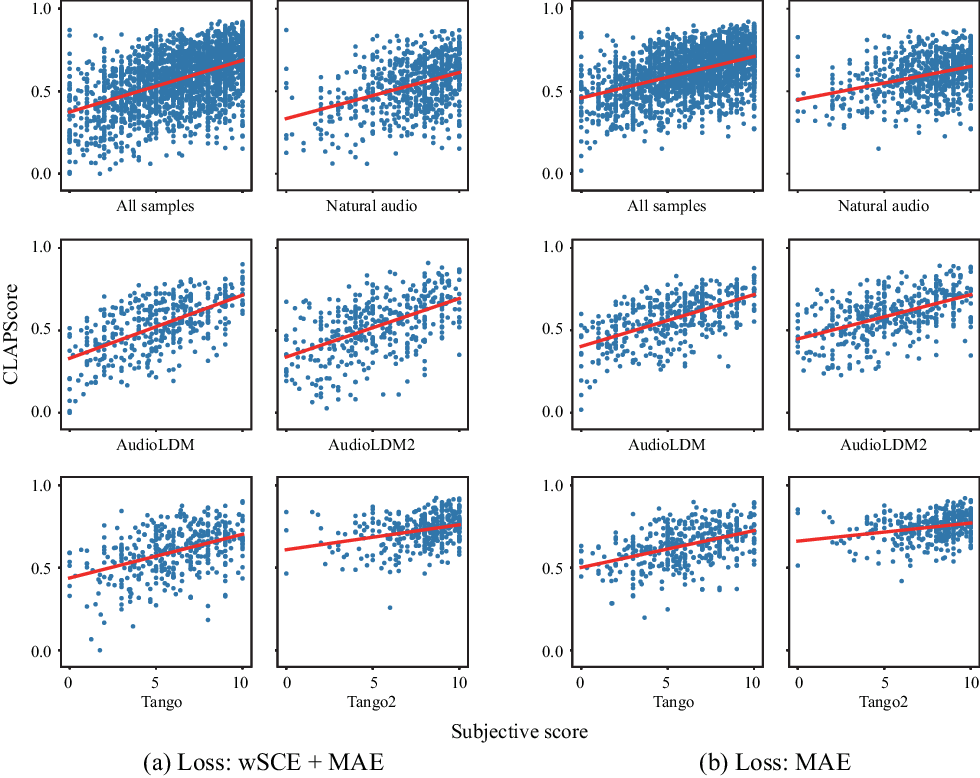}
\caption{Relationship between CLAPScore and subjective evaluation score of Human-CLAP}
\label{fig:scatterplot_human_clap}
\end{figure*}

The overall loss function $L$ is formulated as a linear combination of the wSCE and MSE or MAE losses:
\begin{equation}\label{eq:loss_1}
    L=\lambda_1 L_{\rm wSCE}+\lambda_2L_{\rm reg},
\end{equation}
where $L_{\rm reg}$ is the regression loss, which is $L_{\rm MSE}$ or $L_{\rm MAE}$, $\lambda_1$, and $\lambda_2$ are hyperparameters for controlling the effects of the loss functions.

\section{Experiments}
We compared the relationship between CLAPScores and subjective evaluation scores under different conditions to confirm whether our proposed method is effective in improving the correlation relationship.
\subsection{Experimental Setup}
We fine-tuned LAION CLAP~\cite{laionclap2023} using the subjective evaluation score collected in Section~\ref{sec:clap_subj}.
The learning rate was set to $10^{-5}$ and AdamW~\cite{loshchilov2018decoupled} was used for optimization, and the batch size was 8.
We trained the model for 50 epochs and selected the best model that had the lowest validation loss.
The hyperparameters of $L$ (Eq.~(\ref{eq:loss_1})) were emperically determined to be $[\lambda_1, \lambda_2]=[0.1, 1]$ after multiple rounds of hyperparameter tuning.
We used RoBERTa~\cite{DBLP:journals/corr/abs-1907-11692} and hierarchical token-semantic audio Transformer (HTS-AT)~\cite{chen2022hts} for the text and audio encoders, respectively.
We split the train data collected in Section~\ref{sec:clap_subj} into 1,925 and 458 audio--text pairs for model training and validation.
%



We evaluated the correlation between CLAPScore and subjective evaluation scores using 2,405 audio--text pairs collected from the AudioCaps~\cite{kim2019audiocaps} test set in Section~\ref{sec:clap_subj}.
We used SRCC and LCC to calculate the correlation between CLAPScore and subjective score, and used MSE to evaluate the error between CLAPScore and subjective score.
LAION CLAP~\cite{laionclap2023} and MS CLAP~\cite{elizalde2023clap} are used as the comparison methods.
\begin{table}[tb]
\centering
\setlength{\tabcolsep}{1mm}
\caption{SRCC of CLAPScore and subjective evaluation scores on natural and synthesized audio}
\label{tab:eval-natural-syn-eusipco}
\resizebox{\linewidth}{!}{%
\begin{tabular}{l|c|ccccc}
\Hline
Model & Natural & \multicolumn{5}{c}{Synthesized}\\
\hhline{~|~|-----}
 & & AudioLDM & AudioLDM 2 & Tango & Tango 2 & All \\
\hline
\textbf{Human-CLAP (ours)} & & & & & &\\
\quad wSCE + MSE & 0.331  & 0.549 & 0.487 & 0.385 & 0.275  & 0.548 \\
\quad wSCE + MAE & \textbf{0.358} & \textbf{0.589} & \textbf{0.538} & 0.425 & \textbf{0.310}  & \textbf{0.583} \\
\quad wSCE & 0.303 & 0.485 & 0.465 & 0.326 & 0.230 & 0.488 \\
\quad MSE & 0.338 & 0.513 & 0.485 & 0.354 & 0.278  & 0.527 \\
\quad MAE & 0.333  & 0.567 & 0.513 & \textbf{0.448} & 0.282 & 0.574 \\
\hline
\textbf{Baseline} & & & & & & \\
\quad LAION CLAP & 0.201 & 0.288 & 0.332 & 0.178 & 0.144 & 0.338 \\
\Hline
\end{tabular}
}
\end{table}
\subsection{Results}
\textbf{Overall results}.
Table~\ref{tab:eval-proposed-eusipco} shows the results.
The proposed method improved the SRCC by 0.15 or more compared with baseline models.
In particular, the model fine-tuned with wSCE + MAE achieved the best scores across almost all metrics.
From these results, we confirmed that our proposed method is effective in improving the correlation between CLAPScore and subjective evaluation scores.

Figure~\ref{fig:scatterplot_human_clap}(a) shows the relationship between CLAPScore and subjective evaluation scores of the best-performing model trained with wSCE + MAE.
Compared with the CLAPScore before fine-tuning in Fig. 1, it can be confirmed that the scores are predicted consistently regardless of whether the target subjective evaluation scores are high or low.
These results demonstrate that fine-tuning the CLAP model with the proposed loss function can predict scores that are closely aligned with human subjective evaluations.

Figure~\ref{fig:scatterplot_human_clap}(b) shows the relationship between CLAPScore fine-tuned with only MAE and human subjective score.
As shown in the figure, it can be confirmed that the model trained using only MAE assigns a relatively higher CLAPScore than the model fine-tuned with wSCE + MAE.
This indicates that the model fine-tuned with only MAE struggles to assign low CLAPScores to low subjective scores, particularly those close to 0.
These results suggest that the Human-CLAP fine-tuned with wSCE has more potential to score low-relevant audio-text pairs and to assign low scores to low subjective scores.

\begin{table}[tb]
\centering
\setlength{\tabcolsep}{1mm}
\caption{SRCC depending on the subjective evaluation score. 
Nat. and Syn. denote natural and synthesized audio, respectively.
}
\label{tab:eval-low-high-eusipco}
\resizebox{\linewidth}{!}{%
\begin{tabular}{l|ccc|ccc}
\Hline
 Model & \multicolumn{3}{c|}{Sub. score $\leq 5$}  & \multicolumn{3}{c}{Sub. score $>5$} \\
\multicolumn{1}{c|}{} & All  $\uparrow$ & Nat. $\uparrow$ & Syn. $\uparrow$ & All $\uparrow$ & Nat. $\uparrow$ & Syn. $\uparrow$\\
\hline
\textbf{Human-CLAP (ours)} & & & & & &\\
\quad wSCE + MSE & 0.327 & 0.246 & 0.356 & 0.275 & 0.205 & 0.382 \\
\quad wSCE + MAE & \textbf{0.357} & \textbf{0.287} & \textbf{0.392} & \textbf{0.280} & \textbf{0.229} & \textbf{0.408} \\
\quad wSCE & 0.281 & 0.156 & 0.322 & 0.213 & 0.184 &  0.306  \\
\quad MSE & 0.327 & 0.269 & 0.348 & 0.272 & 0.207 & 0.368   \\
\quad MAE & 0.336  & 0.262 & 0.367 & 0.277 & 0.208 & 0.406  \\
\hline
\textbf{Baseline} & & & & & & \\
\quad LAION CLAP & 0.205 & 0.072 & 0.244 & 0.162 & 0.120 & 0.213 \\
\Hline
\end{tabular}
}
\end{table}
\newpage
\textbf{Results on natural and synthesized audio}.
We divided the test data into natural and synthesized audio and calculated the SRCC between CLAPScore and subjective score.
For the synthesized audio samples, we further calculated the SRCC for each TTA model.
Table~\ref{tab:eval-natural-syn-eusipco} shows the results.
We confirmed that our model has successfully improved the SRCC for all synthesized audio samples.
This result suggests that the CLAPScore obtained using Human-CLAP evaluates synthesized audio samples in TTA tasks more closely to human evaluation, showing its effectiveness for the evaluation in TTA tasks.

\textbf{Results depending on the values of subjective scores}.
We also divided the test data, depending on the values of subjective evaluation scores, to assess the effectiveness of Human-CLAP in evaluating highly related or less related audio--text pairs.
Table~\ref{tab:eval-low-high-eusipco} shows the results.
An increase in SRCC was observed on both high-score and low-score data.
It is noteworthy that there was a significant improvement in SRCC for natural audio, which was notably low in conventional CLAP.
This result indicates that the Human-CLAP can be more useful for the evaluation of natural audio compared with the conventional CLAP.
\section{Conclusion}
In this paper, we investigated the correlation between subjective evaluation scores and CLAPScore, which uses the embedding space of CLAP.
Results showed that the CLAPScore had a low correlation with subjective evaluation scores.
We also proposed Human-CLAP, a human-perception-based CLAP, fine-tuned by using a small amount of subjective evaluation scores.
CLAPScore obtained using Human-CLAP indicated a higher SRCC than the conventional CLAP models by 0.17 or higher.
In particular, the model fine-tuned with wSCE + MAE obtained the best scores across almost all metrics.
In addition, for only synthesized audio, an improvement of approximately 0.25 in SRCC was confirmed compared with the conventional CLAP model.
We confirmed that Human-CLAP can achieve higher SRCC on both natural and synthesized audio samples than conventional CLAP.
In the future, we will collect subjective evaluation scores for other audio datasets and investigate the performance of Human-CLAP for out-of-domain data.


\section*{Acknowledgment}
The work was supported by JSPS KAKENHI Grant Number 24K23880, 25K21221, ROIS NII Open Collaborative Research 2025-(251S4-22735), JST Moonshot Grant Number JPMJMS2237.










\printbibliography

\end{document}